\documentclass[journal]{IEEEtran}
\usepackage{amssymb}
\usepackage{amsfonts}
\usepackage{booktabs}
\usepackage[pdftex]{graphicx}
\usepackage[cmex10]{amsmath}
\usepackage{algorithm}
\usepackage{algorithmic}
\usepackage{array}
\usepackage[caption=false,font=footnotesize]{subfig}
\usepackage{colortbl}

\begin{document}
\title{Distributed Interference Alignment with Low Overhead}
\author{Yanjun~Ma,
        Jiandong~Li,~\IEEEmembership{Senior~Member,~IEEE}, and Rui
        Chen~\IEEEmembership{Student~Member,~IEEE}
\thanks{This work was supported in part by the National Science Fund for Distinguished
Young Scholars under Grant 60725105, by the National Basic Research Program of China
 under Grant 2009CB320404, by the Program for Changjiang Scholars and Innovative
 Research Team in University under Grant IRT0852, by the Key Project of Chinese Ministry of Education under Grant
  107103, and by the 111 Project under Grant B08038.}
\thanks{The authors are with the State Key Laboratory of Integrated
Service Networks, Xidian University, Xi'an 710071, China
 (e-mail: \{yjm, jdli \}@mail.xidian.edu.cn), rchenxidian@gmail.com.}}
\maketitle

\begin{abstract}
Based on closed-form interference alignment (IA) solutions, a low overhead
distributed interference alignment (LOIA) scheme is proposed in this letter for
the $K$-user single-input single-output (SISO) interference channel, and
extension to multiple antenna scenario is also considered. Compared with the
iterative interference alignment (IIA) algorithm proposed by Gomadam \emph{et
al.}, the overhead of our LOIA scheme is greatly reduced. Simulation results
show that the IIA algorithm is strictly suboptimal compared with our LOIA
algorithm in the \emph{overhead-limited} scenario.
\end{abstract}

\IEEEpeerreviewmaketitle

\section{Introduction}
\IEEEPARstart{I}{nterference} alignment has attracted much attention in recent
years, which was introduced by Maddah-Ali \emph{et al}. \cite{ref1}-\cite{ref2}
for the multiple-input multiple-output (MIMO) X channel and subsequently by
Cadambe and Jafar \cite{ref3} in the context of the $K$-user interference
channel (IC). The basic idea of IA is to align multiple interfering signals at
each receiver in order to reduce the effective interference. It was shown in
\cite{ref3} that up to $K/2$ total degrees-of-freedom (DoF) is achievable.

While the huge potential benefits can be achieved via IA, several challenges
must be overcome before these benefits translated into practice. One key issue
is the assumption of global channel state information (GCSI) at each node which
is assumed in most papers. While a node may acquire CSI for its own channels,
it is much harder to learn the channels between other pairs of nodes with which
this node is not directly associated. To eliminate the GCSI assumption, an
iterative interference alignment (IIA) algorithm was proposed in \cite{ref4}
based on channel reciprocity to align interference in a distributed fashion.
However, the overhead induced by long iterations can overwhelm the gain
achieved by IA \cite{ref5}.

In this letter, based on the closed-form IA solutions in \cite{ref3}, a low
overhead distributed interference alignment (LOIA) scheme is proposed for the
$K$-user single-input single-out (SISO) IC, and extension to multiple antenna
scenario is also considered. The overhead of our LOIA scheme is greatly reduced
compared with the IIA algorithm. Simulation results show that the IIA algorithm
is strictly suboptimal compared with our LOIA algorithm in the overhead-limited
scenario.

We have noticed that concurrent work was developed in \cite{ref9} based on the
close-form solutions in \cite{ref6}. However, their scheme is specially
designed for FDD scenario where much overhead and delay are incurred while our
scheme is based on the channel reciprocity in TDD scenario. It is pointed out
in \cite{ref5}, \cite{ref10} that IA is impractical when applied to many users
in the IC simultaneously. So, we are more interested in the 3-user IC.

\section{System Model and Preliminaries}
Consider the $K$-user time-varying or frequency-selective IC consisting of $K$
transmitter - receiver pairs. Each node is equipped with one antenna (multiple
antenna scenario is considered later). The received signal at receiver $k$ is
given by
\begin{equation*}
{Y}^{[k]} = \sum^K_{j=1} \textbf{H}^{[kj]} {X}^{[j]} + {Z}^{[k]},
\end{equation*}
where ${X}^{[j]}$ is an $M \times 1 $ vector that represents the transmitted
signal of user $j$ and $M$ denotes the number of symbol extension of the
channel in time or frequency. ${Z}^{[k]}$ is a zero mean additive white
Gaussian noise. $\textbf{H}^{[kj]}$ is the diagonal channel matrix between
transmitter $j$ and receiver $k$. Let $d^{[k]}$ be the independent data streams
sent from transmitter $k$ to receiver $k$. Transmitter $k$ sends a signal
vector ${S}^{[k]}$ to its intended receiver $k$ by using a precoding matrix
$\textbf{V}^{[k]}$, i.e.,
\begin{equation*}
{X}^{[k]} = \textbf{V}^{[k]} {S}^{[k]}.
\end{equation*}
Receiver $k$ estimates the transmitted data vector ${S}^{[k]}$ by using a
receive beamforming matrix $\textbf{U}^{[k]}$, i.e.,
\begin{equation*}
\tilde{{S}}^{[k]} = (\textbf{U}^{[k]})^{\dag} {Y}^{[k]},
\end{equation*}
where  $(\cdot)^{\dag}$ stands for conjugate transpose.
\subsection{A Brief Review of IIA Algorithm}
\begin{figure}[!tr]
\centering
\includegraphics[width=9cm]{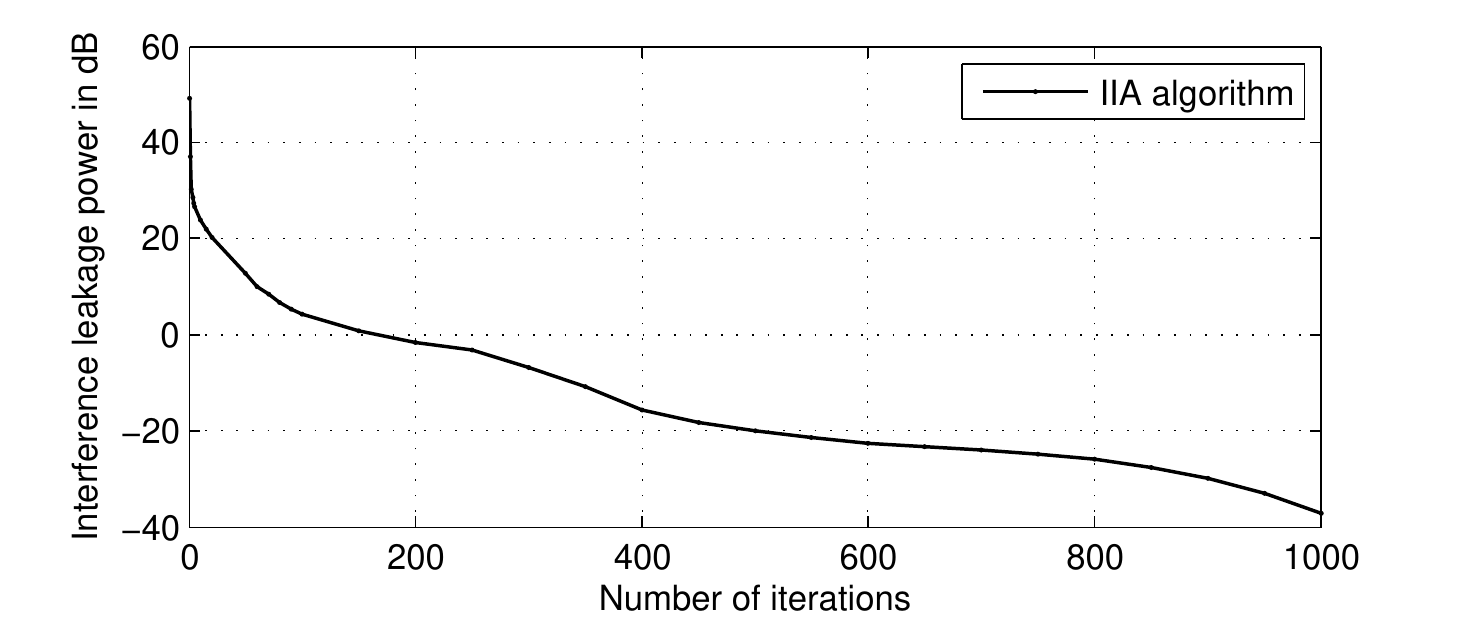}
\caption{Interference leakage power per user of the IIA algorithm (algorithm 1
in \cite{ref4}) in the 3-user MIMO IC where each node is equipped with 2
antennas and one stream is transmitted per user-pair. The transmit power of
each node is 40dB while the noise power is normalized to 0dB.} \label{fig1}
\end{figure}

The IIA algorithm was proposed in \cite{ref4} by utilizing the reciprocity of
wireless channel. Fig.~\ref{fig1} shows the relationship between the number of
iterations and the interference leakage power per user. We can see that long
iterations are needed before data transmission began. The overhead induced by
the iterations can overwhelm the gain achieved by IA \cite{ref5}.

\section{Proposed LOIA Algorithm }
In this section, a low overhead distributed interference alignment (LOIA)
scheme is proposed based on the close-form solutions in \cite{ref3}.
\subsection{LOIA Algorithm for the $K$-User SISO IC}
\begin{figure}[!tr]
\centering
\includegraphics[width=9cm]{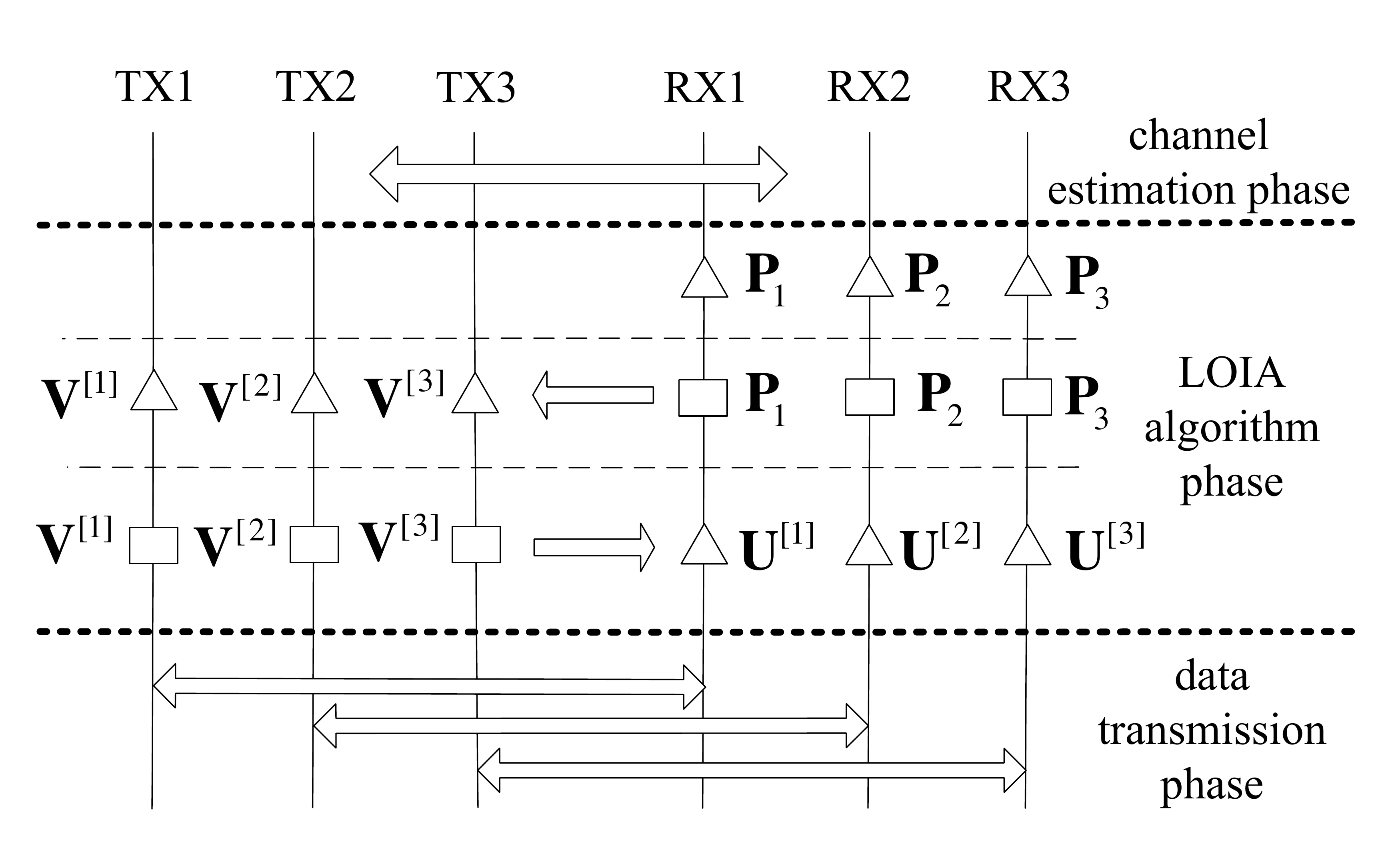}
\caption{The process of the LOIA algorithm for the 3-user IC, where the
triangle marks mean that some calculations are performed on this node and the
quadrangle marks mean that some training sequences are sent from this node
along some vectors.} \label{fig2}
\end{figure}

We describe the LOIA algorithm for the 3-user SISO IC in detail. Extension to
the $K$-user SISO IC can be realized based on Appendix III in \cite{ref3}. For
a limited space, we will not repeat it here. We show that $\{ d^{[1]}, d^{[2]},
d^{[3]} \} = \{ n+1, n, n \}$ data streams can be sent simultaneously over a
$2n+1$ symbol extension in a distributed fashion. The process is divided into 3
phases and is shown in Fig.~\ref{fig2}.
\\\textbf{Phase I}:  Channel estimation phase.
\\\textbf{Phase II}: LOIA algorithm phase.
\\\textbf{1)} At RX1, RX2, and RX3, let
\begin{equation*}
\textbf{P}_1 = \textbf{H}^{[12]} (\textbf{H}^{[13]})^{-1},
\end{equation*}
\begin{equation*}
\textbf{P}_2 = \textbf{H}^{[23]} (\textbf{H}^{[21]})^{-1},
\end{equation*}
\begin{equation*}
\textbf{P}_3 = \textbf{H}^{[31]} (\textbf{H}^{[32]})^{-1},
\end{equation*}
respectively.
\\\textbf{2)} RXs send some training sequences along the
vectors in $\textbf{P}_1$, $\textbf{P}_2$, and $\textbf{P}_3$ respectively.
Then $\textbf{P}_1$, $\textbf{P}_2$, and $\textbf{P}_3$ can be estimated in all
TXs.
\\\textbf{3)} Let $\textbf{w}=[1,1,\dots,1]^T$ be a $(2n+1) \times 1$ column vector,
which is a predefined vector and is known to all TXs. All TXs calculate
\begin{equation*}
\textbf{T} = \textbf{P}_1 \textbf{P}_2 \textbf{P}_3.
\end{equation*}
At TX1, let
\begin{equation*}
\textbf{V}^{[1]} = [\textbf{w}~~\textbf{Tw}~~\textbf{T}^2
\textbf{w}~~\dots~~\textbf{T}^n \textbf{w} ].
\end{equation*}
At TX2, calculate $\textbf{C} = [ \textbf{w}~~\textbf{Tw}~~\textbf{T}^2
\textbf{w}~~\dots~~\textbf{T}^{n-1} \textbf{w}]$, and let
\begin{equation*}
\textbf{V}^{[2]} = \textbf{P}_3 \textbf{C}.
\end{equation*}
At TX3, calculate $\textbf{B} = [\textbf{Tw}~~\textbf{T}^2
\textbf{w}~~\dots~~\textbf{T}^n \textbf{w}]$, and let
\begin{equation*}
\textbf{V}^{[3]} = \textbf{P}_2^{-1} \textbf{B}.
\end{equation*}
Then IA is achieved at all RXs, as $\textbf{V}^{[1]}$, $\textbf{B}$,
$\textbf{C}$, and $\textbf{T}$ satisfy
\begin{equation} \label{eq31}
\textbf{B} = \textbf{T} \textbf{C},
\end{equation}
\begin{equation} \label{eq32}
\textbf{B} \prec \textbf{V}^{[1]},
\end{equation}
\begin{equation} \label{eq33}
\textbf{C} \prec \textbf{V}^{[1]},
\end{equation}
where $\textbf{P} \prec \textbf{Q} $, means that the set of column vectors of
matrix $\textbf{P}$ is a subset of the set of column vectors of matrix
$\textbf{Q}$. It is shown in \cite{ref3} that (\ref{eq31})-(\ref{eq33}) are
equivalent to the IA conditions (\ref{eq10})-(\ref{eq12}).
\begin{equation} \label{eq10}
\textbf{H}^{[12]} \textbf{V}^{[2]} = \textbf{H}^{[13]} \textbf{V}^{[3]},
\end{equation}
\begin{equation} \label{eq11}
\textbf{H}^{[23]} \textbf{V}^{[3]} \prec \textbf{H}^{[21]} \textbf{V}^{[1]},
\end{equation}
\begin{equation} \label{eq12}
\textbf{H}^{[32]} \textbf{V}^{[2]} \prec \textbf{H}^{[31]} \textbf{V}^{[1]}.
\end{equation}
\\\textbf{4)} TXs send some training sequences along vectors in
$\textbf{V}^{[j]}$, $j=1,2,3$. RXs estimate $\textbf{V}^{[j]}$ and let
\begin{equation*}
\textbf{U}^{[1]} = \text{null} ( \textbf{H}^{[12]} \textbf{V}^{[2]} ),
\end{equation*}
\begin{equation*}
\textbf{U}^{[2]} = \text{null} ( \textbf{H}^{[21]} \textbf{V}^{[1]} ),
\end{equation*}
\begin{equation*}
\textbf{U}^{[3]} = \text{null} ( \textbf{H}^{[31]} \textbf{V}^{[1]} ),
\end{equation*}
at RX1, RX2, and RX3 respectively, where $\text{null}(\cdot)$ denotes an
orthonormal basis for the null space of a matrix.
\\\textbf{Phase III}: Data transmission phase.

\emph{Remark: } Let $\textbf{U}^{[j]}$ be the transmit precoding matrix at
$j$-th RX, and let $\textbf{V}^{[j]}$ be the receive beamforming matrix at
$j$-th TX. By using the reciprocity of alignment \cite{ref4}, IA is also
achieved for the reverse transmission.

\subsection{Extension to Multiple Antenna Scenario}
Based on the close-form solution in Appendix IV in \cite{ref3}, our scheme can
be extended to multiple symmetric antenna scenario in the 3-user MIMO IC
without symbol extension in time or frequency. The process is also shown in
Fig.~\ref{fig2}.

\begin{figure}[!tr]
\centering
\includegraphics[width=9cm]{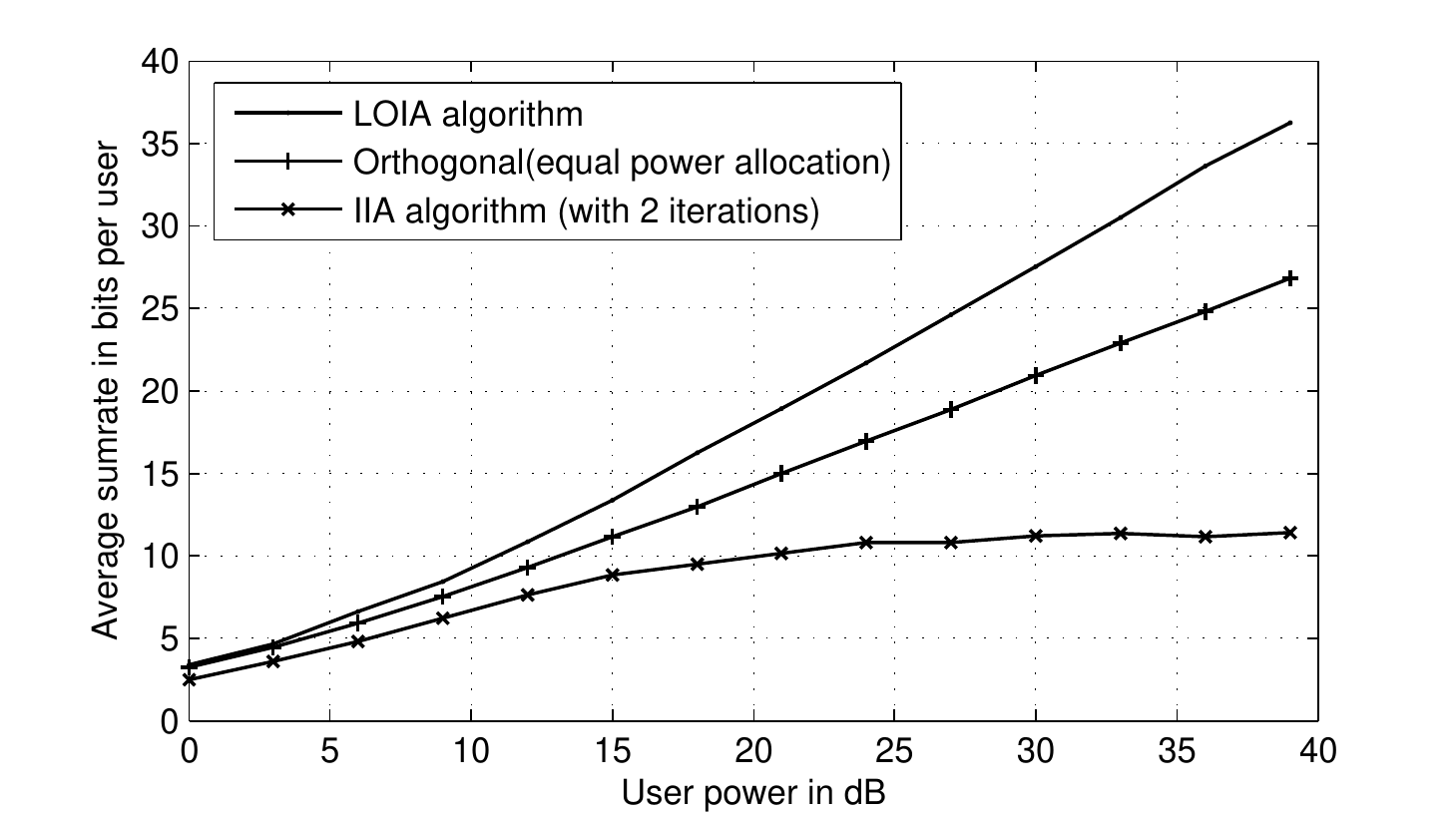}
\caption{Comparison of the LOIA algorithm with the IIA algorithm (algorithm 1
in \cite{ref4}) for the $3$-user MIMO IC in the overhead-limited scenario,
where each node is equipped with 2 antennas and the number of iterations of the
LOIA algorithm and the IIA algorithms equals 2.} \label{fig3}
\end{figure}

We assume each node is equipped with even number of antennas (odd antenna
configuration for the 3-user MIMO IC can be deduced based on Appendix V in
\cite{ref3}). The process is also divided into 3 phases.
\\\textbf{Phase I}:  Channel estimation phase.
\\\textbf{Phase II}: LOIA algorithm phase.
\\\textbf{1)} At RX1, RX2, and RX3, let
\begin{equation*}
\textbf{P}_1 = (\textbf{H}^{[12]})^{-1} \textbf{H}^{[13]},
\end{equation*}
\begin{equation*}
\textbf{P}_2 = (\textbf{H}^{[23]})^{-1} \textbf{H}^{[21]},
\end{equation*}
\begin{equation*}
\textbf{P}_3 = (\textbf{H}^{[31]})^{-1} \textbf{H}^{[32]},
\end{equation*}
respectively.
\\\textbf{2)} RXs send some training sequences along the
vectors in $\textbf{P}_1$, $\textbf{P}_2$, and $\textbf{P}_3$ respectively.
Then $\textbf{P}_1$, $\textbf{P}_2$, and $\textbf{P}_3$ can be estimated in all
TXs.
\\\textbf{3)} Let $\textbf{T}=\textbf{P}_3 \textbf{P}_1 \textbf{P}_2$, and let
\begin{equation}\label{eq20}
\textbf{T} \textbf{V}^{[1]} = \textbf{V}^{[1]}
\end{equation}
at TX1. Then choose
\begin{equation*}
\textbf{V}^{[1]} = [ e_1, \dots, e_{M/2} ]
\end{equation*}
at TX1, where $e_1,~e_2, \dots, e_M$ are the $M$ eigenvectors of $\textbf{T}$.
At TX2 and TX3, they can calculate $\textbf{V}^{[1]}$ as above, and let
\begin{equation}\label{eq21}
\textbf{V}^{[2]} = \textbf{P}_3^{-1} \textbf{V}^{[1]},
\end{equation}
\begin{equation}\label{eq22}
\textbf{V}^{[3]} = \textbf{P}_2 \textbf{V}^{[1]},
\end{equation}
respectively. It is shown in \cite{ref3} that (\ref{eq20})-(\ref{eq22}) are
equivalent to (\ref{eq23})-(\ref{eq25}).
\begin{equation}\label{eq23}
\text{span} ( \textbf{H}^{[12]} \textbf{V}^{[2]} )= \text{span} (
\textbf{H}^{[13]} \textbf{V}^{[3]} ),
\end{equation}
\begin{equation}\label{eq24}
\textbf{H}^{[21]} \textbf{V}^{[1]} = \textbf{H}^{[23]} \textbf{V}^{[3]},
\end{equation}
\begin{equation}\label{eq25}
\textbf{H}^{[31]} \textbf{V}^{[1]} = \textbf{H}^{[32]} \textbf{V}^{[2]},
\end{equation}
where span($\cdot$) denotes the vector space that spanned by the columns of a
matrix. Then IA is achieved at all RXs immediately without other iteration.
\\\textbf{4)} TXs send some training sequences along vectors in
$\textbf{V}^{[j]}$, $j=1,2,3$. RXs estimate $\textbf{V}^{[j]}$ and let
\begin{equation*}
\textbf{U}^{[1]} = \text{null} ( \textbf{H}^{[12]} \textbf{V}^{[2]} ),
\end{equation*}
\begin{equation*}
\textbf{U}^{[2]} = \text{null} ( \textbf{H}^{[21]} \textbf{V}^{[1]} ),
\end{equation*}
\begin{equation*}
\textbf{U}^{[3]} = \text{null} ( \textbf{H}^{[31]} \textbf{V}^{[1]} ),
\end{equation*}
at RX1, RX2, and RX3 respectively. By using the reciprocity of the channel, IA
is achieved at all TXs without other iteration.
\\\textbf{Phase III}: Data transmission phase.

We can see that GCSI is achieved through two training phases. Then precoding
matrices and receive beamforming matrices can be calculated directly.
\subsection{Discussions}
In the $K$-user MIMO IC, When $K>3$, our scheme can be extended to this
scenario whenever there has a close-form solution with symmetric antenna
configuration \cite{ref6}-\cite{ref8}. We leave it for future work for
asymmetric antenna configuration in the $K$-user MIMO IC.

\section{Simulation Results}
The comparison of the IIA algorithm with the LOIA algorithm is presented in
Fig.~\ref{fig3} for the $3$-user MIMO IC, where each node is equipped with 2
antennas. We consider 1 stream allocation per user-pair for the IIA algorithm
and the LOIA algorithm, and the number of iterations of the algorithms equals
two. All channel coefficients are assumed independent and identically
distributed (i.i.d.) zero mean unit variance circularly symmetric complex
Gaussian. Orthogonal scheme is also plotted for comparison, where the sum rate
is calculated assuming equal time sharing for the users, and with power $3P$
per node.

From the simulation results we can see that the IIA algorithm is strictly
suboptimal compared with our LOIA algorithm in the overhead-limited scenario.
The achievable rate of the IIA algorithm even lower than orthogonal scheme in
this scenario.

\section{Conclusion}
Based on the closed-form IA solutions in \cite{ref3}, a LOIA algorithm is
proposed in this letter for the $K$-user SISO interference channel, and
extension to multiple antenna scenario is also considered. Compared with the
IIA algorithm, the overhead of our LOIA scheme is greatly reduced.

\ifCLASSOPTIONcaptionsoff
  \newpage
\fi


\begin{thebibliography}{100}

\bibitem{ref1}
M.~Maddah-Ali, A.~Motahari, and A.~Khandani, ``Signaling over MIMO multi-base
systems: Combination of multi-access and broadcast schemes," \emph{in Proc.
IEEE Int. Symp.Inform.Theory}, 2006.

\bibitem{ref2}
M.~Maddah-Ali, A.~Motahari, and A.~Khandani, ``Communication over MIMO X
channel: Interference alignment, decomposition, and performance analysis,"
\emph{IEEE Trans. Inf. Theory}, vol. 54, no. 8, pp. 3457-3470, Aug. 2008.

\bibitem{ref3}
V.~R.~Cadambe and S.~A.~Jafar, ``Interference alignment and the
degrees of freedom for the K user interference channel," \emph{IEEE
Trans. Inf. Theory}, vol. 54, no. 8, pp. 3425-3441, Aug. 2008.

\bibitem{ref4}
K.~Gomadam, V.~R.~Cadambe, and S.~A.~Jafar, ``A distributed numerical approach
to interference alignment and applications to wireless interference networks,"
\emph{IEEE Trans. Inf. Theory}, vol. 57, no. 6, pp. 3309-3322, Jun. 2011.

\bibitem{ref5}
S.~W.~Peters, and R.~W.~Heath, ``User partitioning for less overhead in MIMO
interference channels," in arXiv:cs.IT/1007.0512v1, Jul. 2010.

\bibitem{ref6}
R.~Tresch, M.~Guillaud, and E.~Riegler, ``On the achievability of interference
alignment in the $K$-user constant MIMO interference channel," \emph{in Proc.
IEEE Workshop on Statistical Signal Processing}, Cardiff, U.K., Sep. 2009, pp.
277-280.

\bibitem{ref7}
C.~Yetis, T.~Gou, S.~Jafar, and A.~Kayran, ``On feasibility of interference
alignment in MIMO interference networks," \emph{IEEE Trans. Signal Processing},
vol. 58, no. 9, pp. 4771-4782, Sep. 2010.

\bibitem{ref8}
G.~Bresler, D.~Cartwright, and D.~Tse, ``Settling the feasibility of
interference alignment for the MIMO interference channel: the symmetric case,"
in arXiv:cs.IT/1104.0888v1, Apr. 2011.

\bibitem{ref9}
S.~Cho, H.~Chae, K.~Huang, D.~Kim, V.~Lau, H.~Seo, and B.~Kim,
``Feedback-topology designs for interference alignment in MIMO interference
channels," in arXiv:cs.IT/1105.5476v1, May. 2011.

\bibitem{ref10}
Y.~Ma, J.~Li, Q.~Liu, and R.~Chen, ``Group Based Interference Alignment,"
\emph{IEEE Commun. Lett.}, vol. 15, no. 4, pp. 383-385, Apr. 2011.
\end{thebibliography}
\end{document}